# Active Particles as Mobile Microelectrodes for Selective Bacteria Electroporation and Transport

Yue Wu[1], Afu Fu[2], and Gilad Yossifon[1]*

[1]Faculty of Mechanical Engineering, Micro- and Nanofluidics Laboratory, Technion – Israel Institute of Technology, Haifa 32000, Israel

[2]Technion Integrated Cancer Center, The Rappaport Faculty of Medicine and Research Institute, Technion-Israel Institute of Technology, Haifa 3109602, Israel

* Corresponding author: yossifon@technion.ac.il

**Abstract**

Self-propelling micromotors are emerging as a promising microscale and nanoscale tool for single-cell analysis. We have recently shown that the field gradients necessary to manipulate matter via dielectrophoresis can be induced at the surface of a polarizable active ("self-propelling") metallo-dielectric Janus particle (JP) under an externally applied electric field, acting essentially as a mobile floating microelectrode. Here, we successfully demonstrated for the first time, that the application of an external electric field can singularly trap and transport bacteria and can selectively electroporate the trapped bacteria. Selective electroporation, enabled by the local intensification of the electric field induced by the JP, was obtained under both continuous alternating current and pulsed signal conditions. This approach is generic and is applicable to bacteria and JP, as well as a wide range of cell types and micromotor designs. Hence, it constitutes an important and novel experimental tool for single-cell analysis and targeted delivery.

**One Sentence Summary:**

This work presents the application of active particles as mobile microelectrodes, where selective bacteria trapping and releasing, transport and electroporation are singularly controlled using an external electric field.

# Introduction

The exciting and interdisciplinary area of self-propelling or "active" particles (also termed micromotors) promises applications in drug delivery(*1*), detoxification(*2*), environmental remediation(*3*), immunosensing(*4*), remote surgery, self-repairing systems, self-motile devices and more(*5*). Motion is achieved by designing particles that can asymmetrically draw and dissipate energy, creating local gradients of force for autonomous propulsion(*6*). With freedom to travel along individual pathlines, i.e., autonomous motion, active particles can cover larger areas and volumes and operate under simpler ambient conditions (i.e., without the necessity for field or chemical gradients) than phoretically driven particles. Active particles can be used for cargo transport (loading and translation on an active particle, i.e., "active cargo carrier") and delivery (release). To date, achieving both propulsion of the active carrier and cargo manipulation (load and release) has only been possible by combining two different mechanisms; self-propulsion can be driven by e.g., electric(*7*), magnetic(*8*) and optical(*9*) external fields(*10*) and even with chemical fuel(*11*), while cargo loading is achieved by e.g., magnetic(*12*), electrostatic(*13*), or biomolecular(*14*) recognition and attraction mechanisms. We recently demonstrated(*15*) a novel unification of carrier propulsion and cargo manipulation, demonstrating that it is possible to singularly control both processes by an applied external electric field. This unification allows for significantly simpler and more robust operation.

We have recently shown(*15*) that the field gradients necessary to manipulate matter via dielectrophoresis (DEP) can be induced at the surface of a polarizable, freely suspended active metallo-dielectric Janus particle (JP) under an externally applied electric field, acting essentially as a mobile floating microelectrode. The importance of this finding is that it offers a label-free method to selectively and dynamically manipulate (load, transport and release) a broad range of organic and inorganic cargo. The DEP force(*16*) results from electric field gradients, generated locally at the carrier particle level, acting on the cargo particle-induced dipole. The DEP force can be either attractive (positive DEP) or repulsive (negative DEP), depending on the relative polarizability of the cargo compared to the medium, which is a function of the material's inherent electrical properties and the frequency of the applied field. Thus, high (low) electric field regions can be used to selectively trap particles exhibiting positive (negative) DEP.

Combining DEP with electrically powered active particle propulsion yields an active carrier that can selectively load, transport and release a broad range of cargos, singularly controlled by an external electric field. Due to their facileness and controllability, we(*17*)(*18*)(*19*) and others(*20*)(*21*)(*6*), have focused on electric fields as a favorable source of energy for active particle propulsion. Importantly, electric fields enable the precise tuning of the induced propulsion forces on active particles in real time(*19*) and avoid issues of finite life and/or non-bio-compatibility of commonly used fuels, such as hydrogen peroxide(*22*)(*23*). Furthermore, simple changes in the frequency of the applied electric fields can give rise to a number of distinct electrokinetic effects that can power locomotion in different ways. Uniquely, under the application of a uniform alternating current (AC) electric field, metallodielectric JPs (where one hemisphere is conducting and the other dielectric) have been shown, by us and others, to respond as active particles(*19*)(*24*), despite the external nature of the applied field. This distinctive feature arises from the propulsive mechanism, either induced-charge electrophoresis(*7*)(*25*) (ICEP) or self-dielectrophoresis (sDEP)(*19*), which is produced on the individual particle level rather than via an externally applied global gradient. It has the advantage of being fuel-free, and mobility is greatest in aqueous electrolytes. Addition of magnetic steering, e.g., magnetizing a ferromagnetic Ni layer coated on half of the JP surface (*26*) and using an external rotating static magnet, enables directed motion and selective trapping of cells.

Electrical, ultrasonic, chemical and mechanical-based lysis of *E.coli* in microfluidic systems have been reviewed by Brown et al. 2008 (*27*). Electrical lysis of cells (i.e. irreversible electroporation) is commonly performed with large electrodes in a batch mode for treatment of a large number of cells, e.g. the microfluidic devices developed by Hugle et al. 2018(*28*), and Vulto et al. 2010 (*29*). Due to the commonly large gap (~6-10 mm) between the electrodes, a relatively high voltage (~280V) is required. Reduction of the applied voltage is possible by electric field intensification, obtained by either a microscale gap between the electrodes (*30*), singular-like geometries (e.g., multiscale electrodes, nanopillar electrodes (*31*), saw-tooth electrodes (*32*)), micropore array (*33*) or nanochannel array-based electroporation (NEP) (*34*), ((*33*, *35*). These local electric field intensification approaches allow for single-cell electroporation (*30*), (*36*) (*36*), which opens up new opportunities in manipulating the genetic, metabolic and synthetic contents of single targeted cells. However, these techniques perform electroporation on the cells that are randomly brought to the regions (i.e., electrodes, pore) of intensified electric field and lack the ability to selectively electroporate specific cells.

To perform selective electroporation of cells, Lundqvist. et al (*37*) used solid carbon fiber microelectrodes, which are mounted on a micro-manipulator, to selectively trap and electroporate single cells. More recently, Nadappuram et al. (*12*) developed a nano-tweezer, which consists of two closely spaced electrodes, with gaps as small as 10-20nm, to extract single molecules/organelles from living cells. Micro-pipette techniques (*38*, *39*) can also perform selective trapping and electroporation. Besides the complexity of fabrication of these electrodes with a nanometer-scale gap, these techniques require high-precision micromanipulators with free access from the outside to the sample and hence, can be considered "invasive" techniques. In contrast, the micromotor can be applied in closed microfluidic chambers, and externally controlled using electric and magnetic fields. Such a design offers inherent intensification of the electric field due to the nanometer gap between the floating metallic patch and the conductive substrate. Together with its directed motion ability, using magnetic steering, our novel approach, using the active particle as a mobile microelectrode, offers a significantly simplified and efficient method of unifying selective trapping and electroporation of cells, singularly controlled via an externally applied electric field.

JP propulsion and cargo manipulation must be operated under AC electric field conditions within the parallel indium tin oxide (ITO)-coated glass slide setup, in order to suppress the generation of the gas bubble products of the faradic reactions on the electrodes (*32*). Although DC pulses are the dominant electroporation mode (*40*)(*41*)(*42*), continuous AC fields have also been reported to cause cell electroporation and lysis(*43*). Herein, we demonstrate that the locally intensified electric field intensity and gradient at the JP level enable both selective collection of *E.coli* as well as effective electroporation of the cells, with a moderate voltage, using either a continuous AC field or a train of pulses for electroporation. A significantly higher electroporation rate was observed for *E.coli* collected by JPs, relative to untrapped *E.coli*, proving that our micromotor-based approach enables targeted electroporation of cells. Hence, in the current study, we extend the application of the JP beyond that of a cargo carrier (*15*) to a platform for local electroporation of selectively trapped cells.

# RESULTS

## Directed motion-based selective trapping and release

To study *E. coli* trapping using a metallo-dielectric Janus sphere in an experimental setup consisting of conductive (ITO-coated glass slides) top and bottom substrates, a *z*-scan (11 planes within 10µm distance from ~2 µm below the substrate) of the JP was performed (see Fig. 1C and E). It was found that *E.coli* were trapped at two locations: 1) between the ITO slide and the metallic side of the JP, 2) at the equator of the polystyrene side of the JP. This is in qualitative agreement with the numerical simulation results indicating that the strongest electric field and field gradients are formed in the inherent small gap between the metallic side of the JP and the wall, with smaller field and field gradients also observed at the equator of the dielectric hemisphere (Fig. 1D and F). As a result, we observed trapping, due to positive DEP, at these locations. It should be noted that in some cases bacteria was also trapped above the JP due to positive DEP resulting from the local large electric field gradients existing at the metallo-dielectric interface (Fig. 1D and F). However, our region of interrogation (i.e. between planes A-A and B-B) was below the top of the JP as we were focused on visualizing only those trapped at or below the equator of the JP

It was also found that the orientations of the *E. coli* trapped in between the metallic side of the JP and the bottom substrate were distinctively different in low and high frequencies. In the low frequency regime (50±20k Hz), the metallic side of the JP was partially screened due to the induced electrical double layer (EDL). Therefore, the electrical field outside the EDL had a non-zero tangential electric field component along which the trapped *E. coli* aligned and seemed to be "standing up" (i.e., with its major axis normal to the ITO substrate) (Fig. 1A). However, at higher frequencies (5M Hz), much beyond the relaxation time of the induced EDL ( $f_{RC} = 1/2\pi\tau \sim 1.6k\ Hz$ (*44*); where $\tau = \lambda a/D$ is the induced charge relaxation, $\lambda = \sqrt{\varepsilon D/\sigma}$ ~40nm is the Debye length, $a = 5\mu m$ is the radius of the JP, $D \sim 2 \cdot 10^{-9}\ m^2/s$ is the diffusion coefficient of the ionic species(*45*), $\sigma \sim 9\ \mu S/cm$ is the measured conductivity of the solution before introduction of the *E. coli*, $\varepsilon \sim 80\varepsilon_0$ is the permittivity of water and $\varepsilon_0$ is the permittivity of the vacuum), there was not sufficient time for charging of the induced EDL, and hence, the electrical field lines were perpendicular to the non-screened metallic hemisphere. In this case, in order to fit into the small gap between the JP and the wall, the *E.coli* had to 'lie down' (i.e., with its major axis parallel to the ITO substrate) (Figure 1B).

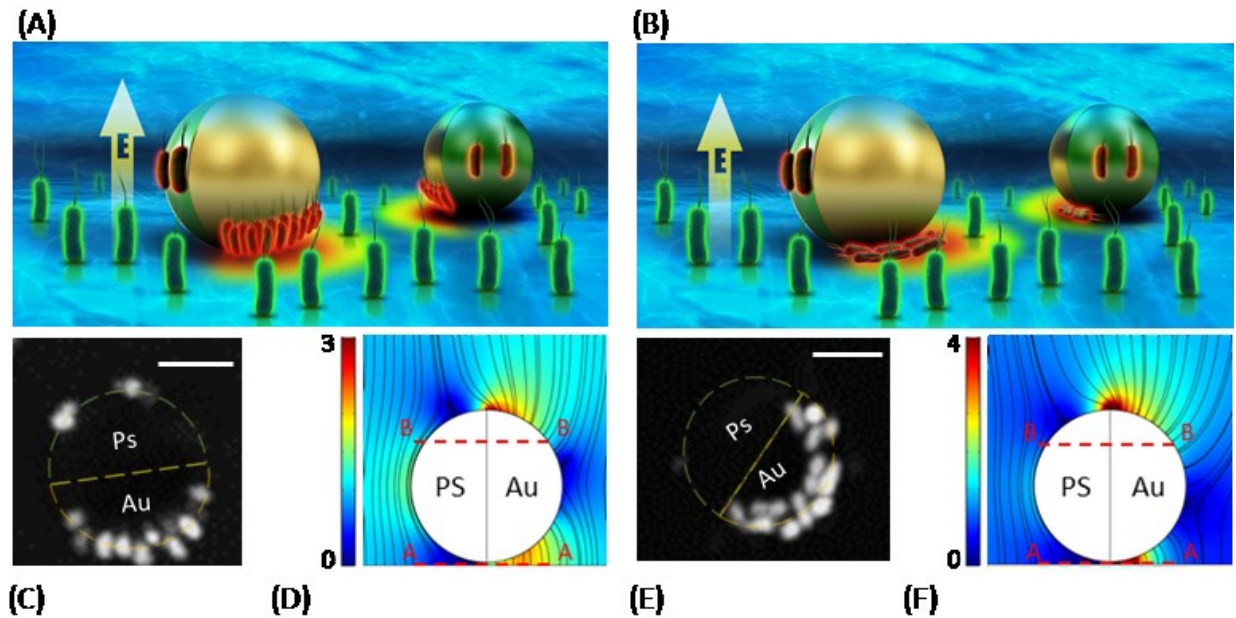

**Fig. 1. Selective bacteria trapping and electroporation.** Schematic illustrations of selective bacteria transport and electroporation, using a Janus particle (JP) as a mobile microelectrode and a continuous AC electric field. The *E.coli* potentially trapped above the JP are not depicted in A and B as we were focused on visualizing only those trapped at or below the equator of the JP. (A, C, D) 50±20k Hz and 15V, (B, E, F) 5M Hz and 15V. (C) and (E) Confocal microscope images obtained by projection of z-scans from plane A-A to B-B (as shown in (D) and (F)). The bacteria that are trapped below the JP exhibit a distinct orientation at low and high electric field frequencies. (D) and (F) are electric field surface plots and streamlines, as obtained from numerical simulations for low ($f/f_{RC}=3$) and high ($f/f_{RC} \gg$) frequencies. Scale bar = 5 μm.

The reason for choosing these two frequencies (50±20kHz, 5 MHz) for studying the continuous AC electroporation was because the JP propulsion velocity at these frequencies approximately vanished (Fig.2C, see also Movie S1 in the Supplementary Materials), which facilitated confocal z-scan imaging at a fixed location. The low frequency regime (50±20k Hz) corresponds to the critical frequency at which the JP reverses direction from ICEP to sDEP motions. This critical frequency depends on the local conductivity of the solution (*46*), which might be affected by cytoplasm coming out from the electroporated cells, and hence varies within a range of frequencies. The exact value of the frequency at which the JP velocity vanishes, which also determines the continuous AC field electroporation conditions, then needs to be determined for each test. The sequential stages of trapping, electroporation and release are described in Fig. 2 A-B (see Movies S2 and S3 in the Supplementary Materials). The JP first collected *E. coli* in the ITO chamber (300 kHz and 10 V, 2 min), moving at a velocity of ~20 µm/s, which is around the maximum velocity achieved in sDEP mode (Fig. 2C). After two minutes of trapping, the electrical field parameters were changed to the desired continuous AC field electroporation conditions. As a result, the outer layer of the trapped *E. coli* became un-trapped (weak DEP force relative to thermal motion) and diffused away. Propidium iodide (PI) uptake by *E. coli* was used as an indication of membrane electroporation(*47*). The electroporation efficiency, defined as the ratio of the number of electroporated cells relative to the total number of trapped cells, was below 10% during the trapping stage, which suggests the ability to carry intact biological matter and electroporate it at a second location.

As the voltage increased, the number of *E. coli* trapped between the JP and the ITO substrate and on the equator of the polystyrene side also increased (Fig. 2(D)), as was expected due to the increased DEP force (*17*). Interestingly, in the low-frequency case (50±20k Hz), the number of trapped *E. coli* increased with time. This was due to the induced-charge electro-osmotic flow (ICEO) (*18*) generated at the metallic hemisphere in the form of jetting, which brought more *E. coli* from the polystyrene side of the JP (see Movie S4 in the Supplementary Materials). In contrast, in the high-frequency case (5M Hz), where there was no electroconvection since the frequency was significantly higher than the RC frequency of the induced-charge, the number of *E. coli* trapped at the JP remained un-changed.

During the release stage (i.e. electric field is turned off), a large number of *E. coli* that were trapped above the JP but could not be visualized during the field operation were released. Hence, in this work, we only studied the electroporation efficiency of the *E. coli* trapped between the JP and ITO substrate as well as those trapped at the equator of the polystyrene side as these could be visualized during the entire process.

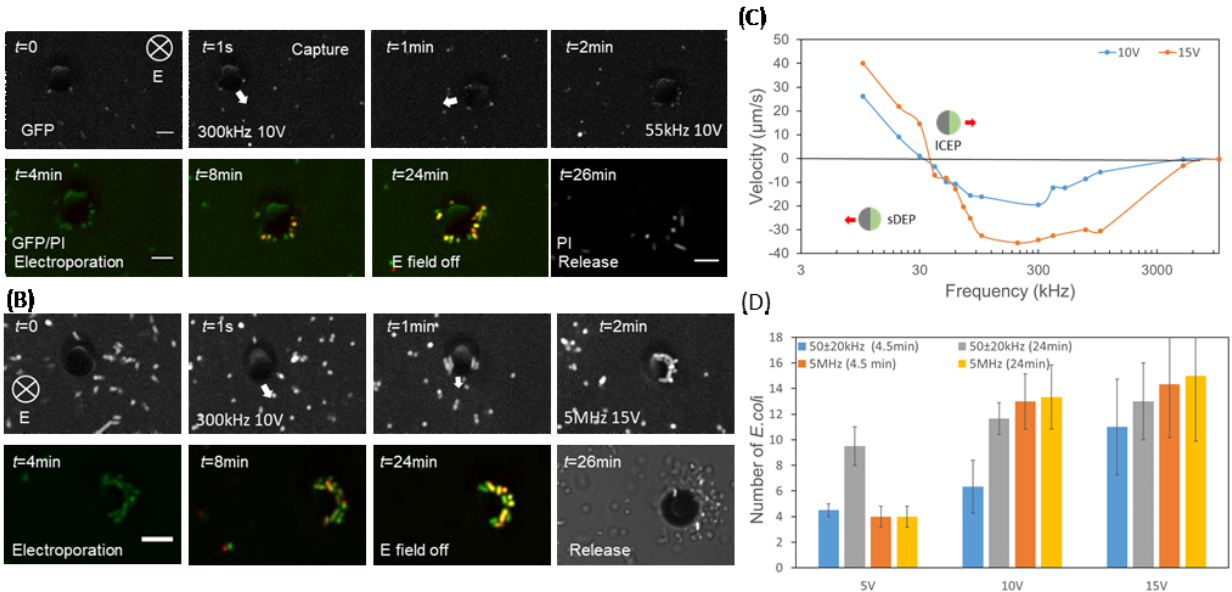

**Fig. 2. Sequential trapping, electroporation and release of *E. coli*.** Trapping was conducted at 300k Hz and 10 V, with a moving Janus particle (JP), while electroporation was performed at either: (A) 50±20k Hz (see Movie S2 in the Supplementary Materials) or (B) 5M Hz (see Movie S3 in the Supplementary Materials), where the JP was immobile, as evident from (C), depicting the JP particle velocity versus frequency (see Movie S1 in the Supplementary Materials). The applied voltage in the electroporation stage varied, but results of the case of 10 V and 15 V are shown here. (D) Number of *E. coli* trapped between the JP and the ITO glass, including at the JP equator, at the low (50±20k Hz) and high (5M Hz) frequency regimes, for different operation times, and varying voltages. Error bars represent standard deviation computed from three independent tests. Scale bar = 5 μm.

**Local and selective electroporation under continuous AC field**

Figure 3 depicts the micrographs of the PI-stained *E. coli* within the ITO chamber under various AC field parameters and operation times. As expected, the PI uptake for trapped *E. coli* increased with the voltage, due to the increased voltage drop across the cell membrane (i.e., transmembrane potential $\Delta\psi_{membr}$), which in turn, resulted in increased electroporation (*48*), in accordance with Schwan's equation (*49*).

PI uptake rate was significantly higher in the trapped versus un-trapped *E. coli*. For example, under low frequency (50±20k Hz, 10 V) (Fig.3), 80% of the trapped *E.coli* were stained at 10 min, whereas, most of the un-trapped *E. coli* were still intact (20% PI uptake). Under high frequency (5M Hz, 10 V) (Fig.4), 45% of the trapped *E. coli* were stained at minute 10, whereas, only ~18% of the un-trapped *E.coli* were stained. However, while at high frequency, PI uptake increased monotonically with increasing voltage, at low frequency, PI uptake at 5 V was higher than at 15 V from minutes 4-17 (Fig.3). This unexpected result might partly stem from the fact that there were multiple layers of *E. coli* trapped at 15V, where the bacteria located at the outer layers were less affected by the intensified electric fields as they were located further away from the center of the JP and because the electric fields were screened by the cells within the inner layers. In addition, the induced electroconvective flow generated by the metallic hemisphere of the JP may have continuously pumped intact *E. coli* from the bulk region from the JP dielectric side. Furthermore, the PI uptake rate of trapped *E. coli* at high frequency was lower than at low frequency. For example, at 5V, up to ~75% of the cells were stained at the high frequency (Fig.4) and 100% at low frequency (Fig.3). This is expected, as the transmembrane potential decreased with increasing frequency (*32*), according to Eqs. 2 and 3 (see Table S1 in the Supplementary Materials).

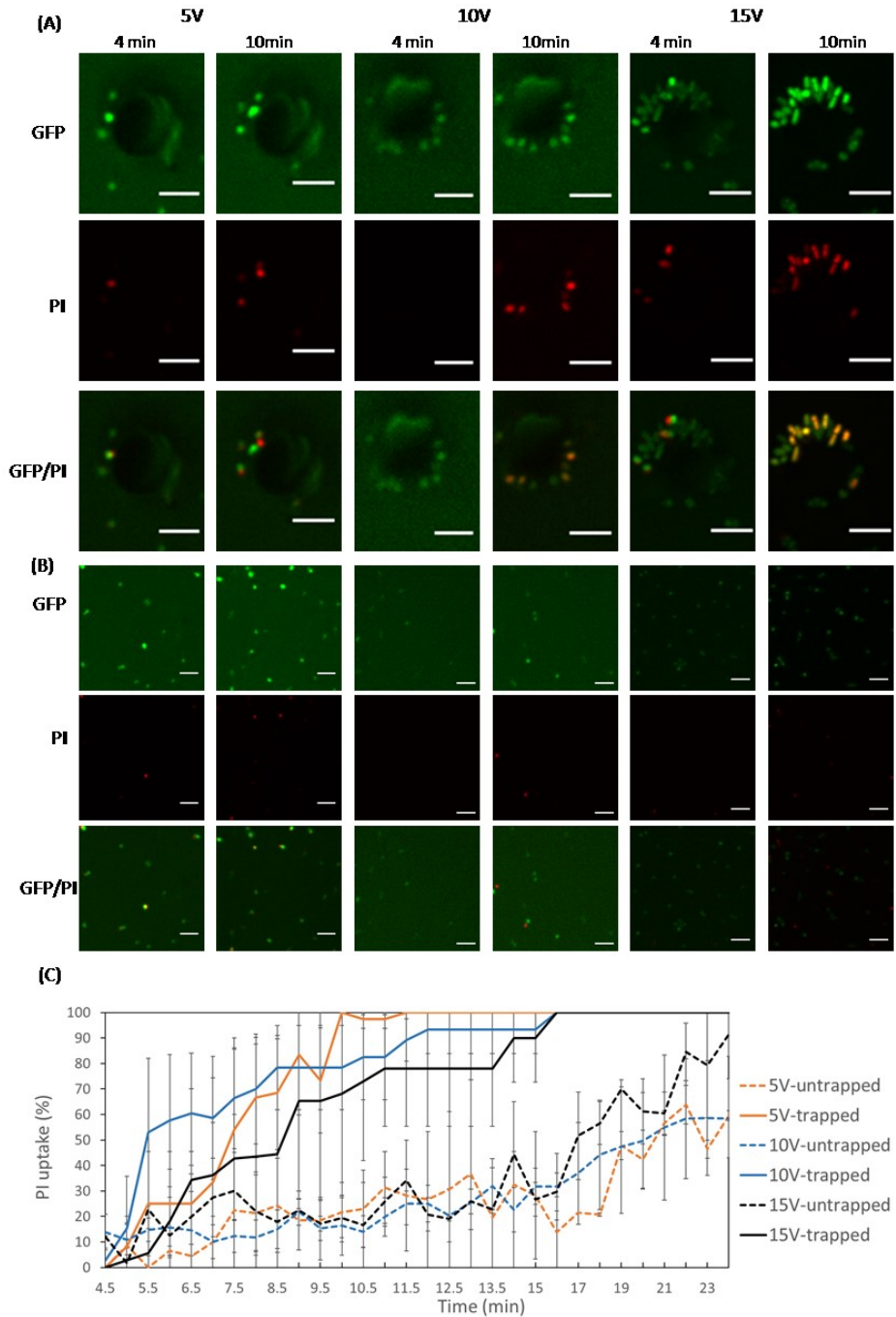

**Fig. 3. Microscope images of PI-stained trapped *versus* non-trapped *E. coli*, at low frequency (50±20k Hz), various applied voltages, and operation times.** PI uptake (red fluorescence) indicates cell electroporation. Microscopic images of (A) trapped *E.coli* and (B) untrapped *E.coli*. (C) Percentage of both trapped and un-trapped PI-stained *E.coli* over time, at various applied voltages (~33k Hz). Error bars represent standard deviation computed from three independent tests. Janus particle of 10µm in diameter was used. Scale bar = 5 µm.

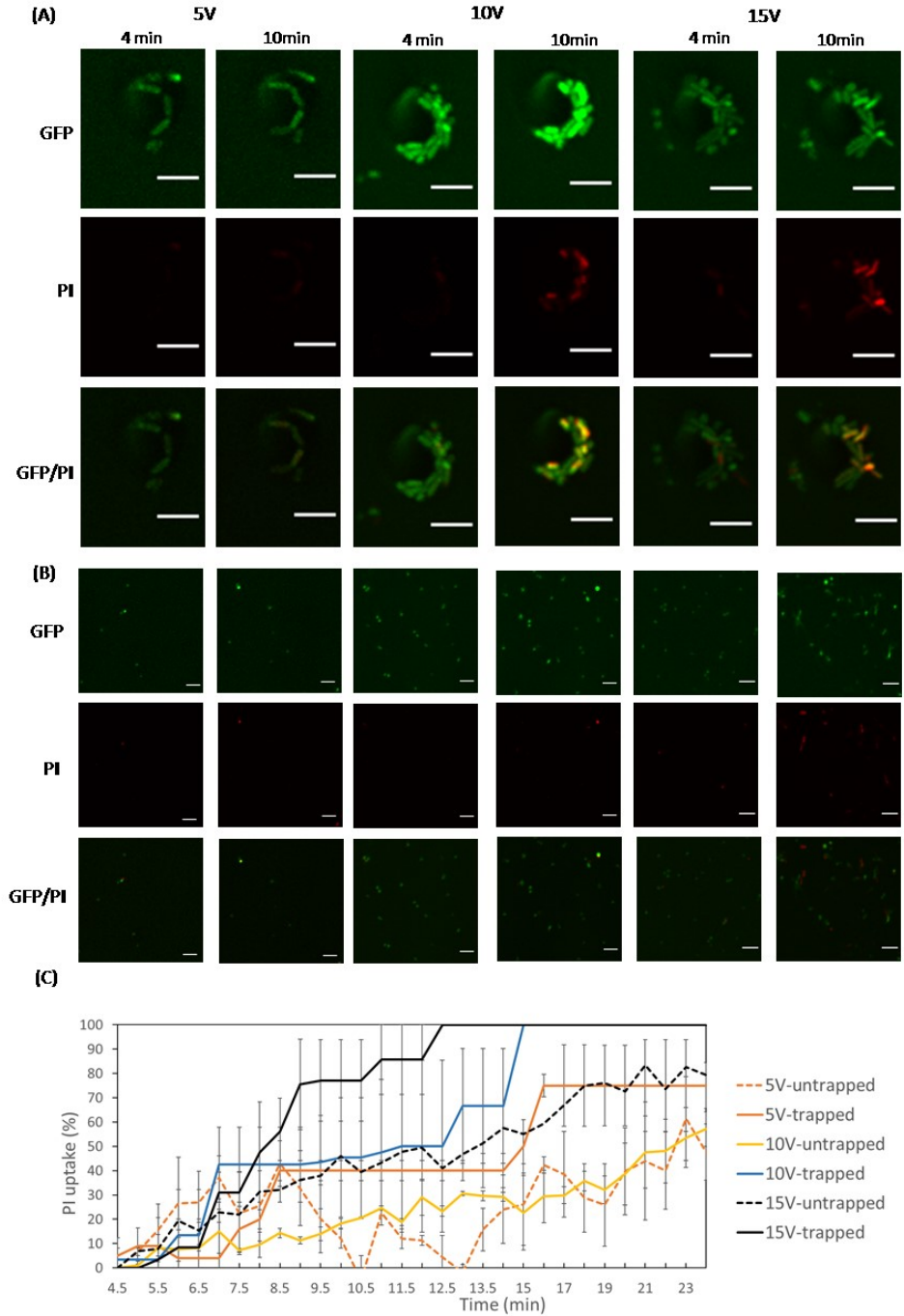

**Fig. 4. Microscope images of PI-stained trapped *versus* non-trapped *E. coli*, at high frequency (5M Hz), various applied voltages, and operation times**. PI uptake (red fluorescence) indicates cell electroporation. (A) Trapped and (B) untrapped *E.coli*. (C) Percentage of both trapped and un-trapped PI-stained *E.coli*, at various applied voltages (5M Hz). Error bars represent standard deviation computed from three independent tests. Janus particle of 10 μm in diameter was used. Scale bar = 5 μm.

**Local and selective electroporation under AC pulses**

In order to suppress continuous electroporation while holding the bacteria, it is preferred to use high frequency, which maintains a lower transmembrane potential (e.g., 5M Hz instead of 50±20k Hz). In this manner, on-demand electroporation can be achieved by combining short pulses (*50*) with a continuous AC signal. Moreover, the electroconvection flow which occurs under the low frequency regime and may adversely influence the cell status (e.g., bringing new bacteria, shearing trapped bacteria etc.), is completely suppressed. When applying an AC pulse train, the JP remained on the substrate in contrast to the case of DC pulse, which seemed to levitate the JP and in this manner, lose the trapped cells. In addition, AC signals tend to reduce electrolysis relative to DC signals (*50*). As shown in Fig. 5, application of a train of 10 AC pulses (see Movie S5 in the Supplementary Materials) yielded a significantly higher percentage of PI-stained cells (100%) compared with a train of 5 (25%) or 1 AC pulse (15%) applied over the same incubation time (4 min). Moreover, after 10 pulses and 4 minutes of incubation, most of the untrapped *E. coli* were still intact (< 25% PI uptake). It should be noted that we made no attempt to optimize the parameters (e.g., duration, peak intensity, interval between pulses etc.) of the pulses and still obtained a clear differentiation in the electroporation response between the conditions at the JP versus the untrapped *E. Coli*. Such parameters can be tuned in the future when reversible or non-reversible electroporation conditions are required.

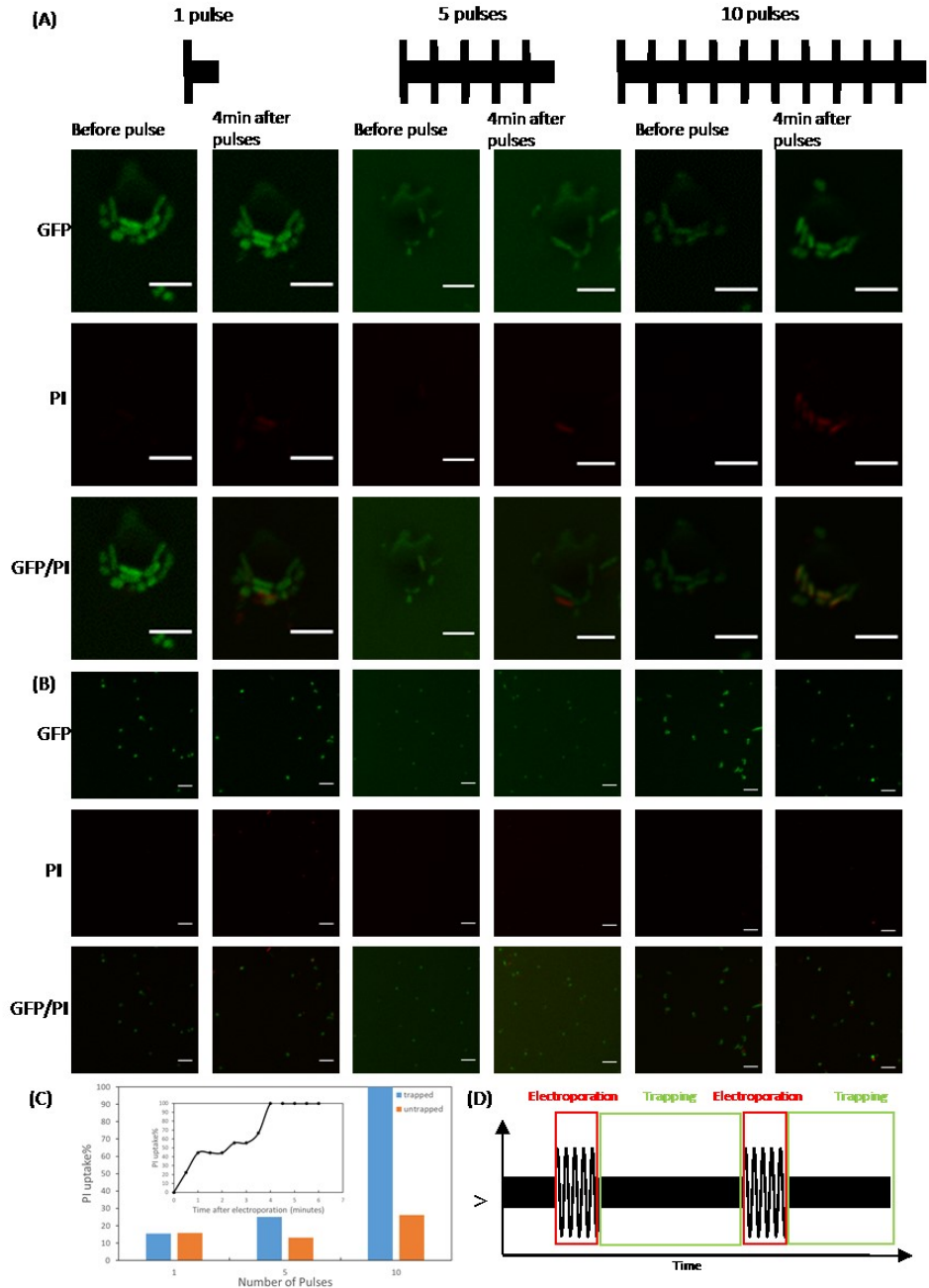

**Fig. 5. Microscope images of PI-stained trapped *versus* non-trapped *E. coli*, under an AC pulse train, various number of pulses, and operation times.** PI uptake (red fluorescence) indicates cell electroporation. (A) Trapped and (B) untrapped *E.coli*. (C) Percentage of both trapped and un-trapped PI-stained *E.coli*, 4 minutes after pulses (frequency: 2M Hz, voltage: 10 V, duration: 0.5ms) Insert: Percentage of PI-stained cells over time, following a train of 10 pulses. (D) Variation of AC voltage and frequency across the experiment chamber and over time. Janus particle of 10 μm in diameter was used. Scale bar = 5 μm.

## Discussion

An electrokinetically driven JP can function as a mobile microelectrode, capable of manipulating cargo via DEP, and serve as a platform for electroporation of cells, due to locally intensified electrical field. In this work, we successfully unified the selective transport, loading and electroporation of biological cargo by simply changing the frequency and amplitude of the applied electric field.

Regarding the trapping capacity of the JP, it was clearly shown that at a frequency of 300k Hz, the number of trapped *E. coli* increased with increasing voltage, as expected, and in agreement with our previous results of JP trapping of polystyrene particles (*17*). Interestingly, the *E. coli* trapped at low frequency versus high frequency showed distinct orientations resulting from the different electric field streamlines obtained under the different frequency regime. This observation was due to the non-spherical (rod) shape of the bacteria, in contrast to the previously studied spherical polystyrene particles. Essentially, at frequencies much lower than the relaxation frequency of the induced electric double layer (EDL), the metallic hemisphere is electrically screened and the electric field lines outside the EDL are mostly tangential to the JP surface, resembling the electrostatic solution around an insulator. At very high frequencies, wherein there isn't sufficient time for the charging of the induced EDL, the metallic hemisphere is not screened and the electric field lines are perpendicular to it, resembling the electrostatic solution around a conductor.

The most important finding of the current study was the ability of the JP to selectively electroporate the trapped cells due to the local electric field intensification, induced by the JP itself, at two locations: 1) under the metallic surface and 2) at the equator of the polystyrene surface. Electroporated cells were stained with PI and their percentage increased with increasing voltages, at all frequencies. Moreover, for the same applied voltage, the PI uptake rate was higher at the lower frequency (50±20k Hz), in agreement with the Schwan's equation for the transmembrane potential. It was found that at the end of the trapping stage (2 minutes), less than 10% of the trapped cells were electroporated, which enabled the collection and transport of intact *E. coli* to a secondary location, where they were then electroporated and further analyzed. Taken together, the JP can be applied to pre-concentrate and electroporate trace amounts of bacteria in samples, which can then be analyzed in a short time period. This application is important in water safety monitoring, health surveillance, and clinical diagnosis, where detection and identification of trace amounts of viable bacterial pathogens is in high demand (*51*, *52*). It is expected that the described biological cargo carrier and targeted electroporation can be used in applications integrating single-cell analysis methods, such as PCR, gene sequencing, fluorescence in situ-hybridization and immunofluorescence staining, where the carrier will selectively pick up a target and transport it to a secondary chamber to be lysed for further analysis of its genomics(*53*), transcriptomics(*54*), proteomics(*55*) or metabolomics(*56*). The selective trapping and single-cell lysis system also enable investigation of cell heterogeneity(*57*).

The observation that an AC pulse train can electroporate cells, while keeping them trapped, opens the opportunity of introducing large molecules and plasmid DNA into the bacteria, which can be of significant utility in gene cloning and research of molecular biology (*58*). The JP can accurately trap the desired number of DNA plasmids to be transfected into the target cell(*59*). There is very limited literature addressing means of controlling the precise number of DNA plasmids to be electroporated into cells (including bacterial). Current transfection methods (cuvette, and micropipette methods(*60*)) roughly control the amount of plasmid being transfected by controlling the bulk concentration of the plasmid.

The fact that the active particle achieved its highest velocity and pDEP force in medium with low conductivities (<0.03S/m), may require exchanging the physiological solution with solution of low ionic conductivity but of similar osmolality, when needed (e.g., mammalian cells). Such exchange of buffer is commonly performed in standard electroporation procedures(*61*). For bacterial and yeast cells, the use of low-conductivity solution is not a limitation. It may also be solved for physiological medium conditions by combining a non-electrokinetic propulsion mechanism (*10*, *11*, *62*), in combination with electrical-based DEP for manipulation (load, release) of the biological cargo and electroporation. Thus, the novel and ground-breaking concept of a unified selective transport and electroporation using the mobile microelectrode-based micromotor described above enables the development of novel microscale and nanoscale tools for single-cell analysis and opens new opportunities for targeted delivery.

## Methods

**Cell culture and preparation of bacteria solution**

*Escherichia coli* (*E. coli*) strain XL1-Blue bacteria were cultured at 37°C and 250 rpm in Luria-Bertani (LB) medium containing 10 g/L tryptone, 5 g/L yeast extract and 10 g/L NaCl. Electorcompetent cells were prepared when the cultures reach an OD600 of 0.5-0.7 using a glycerol method and frozen in a -80 °C freezer. To label bacteria with GFP, a pCDNA3.1-GFP plasmid was transformed to XL1-Blue competent cells by electroporation (1800w, 0.5ms). Cells were cultured in a LB-Agar plate with ampicillin resistance to grow as single colonies. Bacteria were picked up from a colony for experiments using a pipette tip and incubated for 5 min, at room temperature, in 200μL DI water, which contained 33 μg/ml PI, 0.01% (v/v) Tween 20 (Sigma Aldrich) and $7 \cdot 10^{-5}$ M KCl.

**Experimental set-up**

The experimental chamber consisted of a 120 μm-high, silicone reservoir (Grace-Bio), sandwiched between an ITO-coated, 1 mm glass slide (Delta Technologies) and an ITO-coated coverslip (SPI systems), as illustrated in Boymelgreen et al. (*46*) (see Fig.S1 in the Supplementary Materials). Two inlet holes (~1 mm in diameter) were drilled through the top 1 mm ITO slide, surrounded by a silicone reservoir (2 mm in height and 9 mm in diameter) filled with solution, to ensure the chamber remained wet and to enable the addition of the solution with the JPs, bacteria, fluorescent dyes and tracer particles into the channel via manual pumping. The AC electrical forcing was applied using a signal generator (Agilent 33250A) and monitored by an oscilloscope (Tektronix-TPS-2024). An AC pulse signal was applied using a signal generator (TTi TGA 12104 series) with multiple channels. A lab-made switch (Solid State Relays (AQV252G) controlled by Arduino Nano) was used to control the duration and timing of AC pulses. A power amplifier (Falco System) was used to amplify the output signal.

**Selective trapping of *E. coli* by Janus particles and electroporation**

Active collection of *E. coli* was performed in the ITO chamber at 300k Hz and 10 V, for 2 min. In the following 2 min, the electrical field parameters were changed to find the conditions (i.e., zero velocity of the JP) for the electroporation stage during which un-trapped *E. coli* diffuse away. The remaining trapped *E. coli* underwent electroporation using either continuous AC signal for different amplitudes, frequencies and durations, as depicted in Fig. 3A and Fig. 4A, or using an AC pulse train under a continuous AC signal with frequency of 5M Hz. The AC pulse train consisted of electroporation pulses (frequency: 33k Hz, 30 V, duration: 0.5 ms) with 1 s intervals of a continuously applied electric field (frequency: 2M Hz, 10 V) for trapping purposes.

**Microscopy and image analysis**

Trapped and untrapped *E. coli* (Fig. 3A, Fig. 4A, and Fig. 5A) were observed using a Nikon Eclipse Ti-E inverted microscope equipped with a Yokagawa CSU-X1 spinning disk confocal scanner and Andor iXon-897 EMCCD camera. The chamber was placed with the coverslip side down and images were taken using an x60 oil immersion lens. GFP and PI fluorescent dyes were observed with lasers of wavelength 488 nm and 561 nm, respectively. The PI uptake (Fig. 3C) was computed as the ratio between the number of PI-stained trapped *E. coli* and the total number of trapped *E. coli*, after subtraction of the number of PI-stained *E. coli* at minute 4. Here, again the bacteria trapped above the JP were not included.

**Numerical simulations**

The numerical simulation used to qualitatively verify the presence of asymmetric electric field gradients arising from the proximity of a Janus sphere near a conducting wall, was performed in COMSOL™ 5.3. A simple 2D geometry, consisting of a rectangular channel, 50 μm height and 100 μm width, with a 10 μm diameter circle placed 300nm above the substrate, was used to model the experimental setup (*17*). Since the EDLs are thin relative to the radius of the particle ($\lambda/a \ll$ , within the electrolyte we can solve the Laplace equation for the electric potential, $\phi$, in conjunction with the following boundary condition at the metallic side of the JP

$$\sigma \frac{\partial \phi}{\partial n} = i\omega C_{DL}\left(\phi - V_{floating}\right), \quad (1)$$

which describes the oscillatory Ohmic charging of the induced EDL, wherein $V_{floating}$ is the floating potential of the metallic hemisphere of the JP, $n$ is the coordinate in the direction of the normal to the JP surface, and $C_{DL}$ represents the capacitance per unit area of the EDL and may be estimated from the Debye-Huckel theory as $C_{DL} \sim$  . In addition, a floating boundary condition (*25*) was applied on the metallic hemisphere so as to obey total zero charge. An insulation boundary condition was applied on the dielectric hemisphere of the JP, a voltage of 6.25 V was applied at the lower substrate (*y*=0), while the upper wall was grounded, and the edges of the channel were given an insulating boundary condition.

**Calculation of the transmembrane potential**

The following Schwan's equation (*49*) was used for the approximation of the transmembrane potential of *E. coli*:

$$\Delta\psi_{membr} = 1.5aE_{appl}\cos\theta/[1 + (\omega\tau)^2]^{1/2}, \quad (2)$$

$$\tau = aC_{membr}(\rho_{int} + \frac{\rho_{ext}}{2}), \quad (3)$$

where $\theta$ is the angle between the electric field direction and the normal to the cell membrane, $E_{appl}$ is the applied field strength, $f$ is the frequency, $\omega=2\pi f$ is the angular frequency, $a$ is the radius of the cell, $C_{membr}$ (F/cm$^2$) is the capacitance of the membrane, $\rho_{int}$ is the resistivity of the internal fluid, $\rho_{ext}$ is the resistivity of external medium, and $\tau$ is the membrane relaxation time. As an example, for applied voltage difference of 10 V, the calculated transmembrane potential is 0.1 V and 0.002 V, for frequencies of 33k Hz and 5M Hz, respectively (see Table S1 in the Supplementary Materials), demonstrating the decreased electroporation efficiency with increasing frequency.

*RSC Adv.* **8**, 20124–20130 (2018).

29. P. Vulto *et al.*, A microfluidic approach for high efficiency extraction of low molecular weight RNA. *Lab Chip*. **10**, 610–6 (2010).

30. J. Olofsson *et al.*, Single-cell electroporation. *Curr. Opin. Biotechnol.* **14**, 29–34 (2003).

31. C. Xie, Z. Lin, L. Hanson, Y. Cui, B. Cui, Intracellular recording of action potentials by nanopillar electroporation. *Nat. Nanotechnol.* **7**, 185–190 (2012).

32. H. Lu, M. A. Schmidt, K. F. Jensen, A microfluidic electroporation device for cell lysis. *Lab Chip*. **5**, 23 (2005).

33. L. Chang *et al.*, Magnetic tweezers-based 3D microchannel electroporation for high-throughput gene transfection in living cells. *Small*. **11**, 1818–1828 (2015).

34. L. Chang *et al.*, 3D nanochannel electroporation for high-throughput cell transfection with high uniformity and dosage control. *Nanoscale*. **8**, 243–252 (2016).

35. L. Chang *et al.*, Dielectrophoresis-assisted 3D nanoelectroporation for non-viral cell transfection in adoptive immunotherapy. *Lab Chip*. **15**, 3147–3153 (2015).

36. Y. Zhan, J. Wang, N. Bao, C. Lu, Electroporation of Cells in Microfluidic Droplets. *Anal. Chem.* **81**, 2027–2031 (2009).

37. J. A. Lundqvist *et al.*, Altering the biochemical state of individual cultured cells and organelles with ultramicroelectrodes. *Proc. Natl. Acad. Sci. U. S. A.* **95**, 10356–60 (1998).

38. E. N. Tóth *et al.*, Single-cell nanobiopsy reveals compartmentalization of mRNAs within neuronal cells. *J. Biol. Chem.* **293**, 4940–4951 (2018).

39. R. Adam Seger *et al.*, Voltage controlled nano-injection system for single-cell surgery. *Nanoscale*. **4**, 5843 (2012).

40. T. Matos *et al.*, Nucleic acid and protein extraction from electropermeabilized E. coli cells on a microfluidic chip. *Analyst*. **138**, 7347 (2013).

41. S. Homhuan, B. Zhang, F.-S. Sheu, A. A. Bettiol, F. Watt, Single-cell electroporation using proton beam fabricated biochips. *Biomed. Microdevices*. **14**, 533–540 (2012).

42. B. I. Morshed, M. Shams, T. Mussivand, Investigation of Low-Voltage Pulse Parameters on Electroporation and Electrical Lysis Using a Microfluidic Device With Interdigitated Electrodes. *IEEE Trans. Biomed. Eng.* **61**, 871–882 (2014).

43. H. Sedgwick, F. Caron, P. B. Monaghan, W. Kolch, J. M. Cooper, Lab-on-a-chip technologies for proteomic analysis from isolated cells. *J. R. Soc. Interface*. **5 Suppl 2**, S123-30 (2008).

44. A. Boymelgreen, G. Yossifon, Observing Electrokinetic Janus Particle–Channel Wall Interaction Using Microparticle Image Velocimetry. *Langmuir*. **31**, 8243–8250 (2015).

45. D. Ben-Bassat, A. Boymelgreen, G. Yossifon, The influence of flow intensity and field frequency on continuous-flow dielectrophoretic trapping. *J. Colloid Interface Sci.* **442**, 154–161 (2015).

46. A. Boymelgreen, G. Yossifon, T. Miloh, Propulsion of Active Colloids by Self-Induced Field Gradients. *Langmuir*. **32**, 9540–9547 (2016).